# Avoiding a Spanning Cluster in Percolation Models

Y. S. Cho[1], S. Hwang[1], H. J. Herrmann[2], and B. Kahng[1]*

[1]Department of Physics and Astronomy, Seoul National University, Seoul 151-747, Korea
[2]Computational Physics for Engineering Materials, IfB, ETH Zurich, Schafmattstrasse 6, 8093 Zurich, Switzerland

*Correspondence to: bkahng@snu.ac.kr

**When dynamics in a system proceeds under suppressive external bias, the system can undergo an abrupt phase transition, as it occurs for example in the epidemic spreading. Recently, an explosive percolation (EP) model was introduced in line with such phenomena. The order of the EP transition has not been clarified in a unified framework covering low-dimensional systems and the mean-field limit. We introduce a stochastic model, in which a rule for dynamics is designed to avoid the formation of a spanning cluster through competitive selection in Euclidean space. We show by heuristic arguments that, in the thermodynamic limit and depending on a control parameter, the EP transition can be either continuous or discontinuous if $d < d_c$ and is always continuous if $d \geq d_c$, where $d$ is the spatial dimension and $d_c$ the upper critical dimension.**

**One Sentence Summary**: We develop theoretical arguments to determine the order of abrupt percolation transitions in a model that provides a unified framework covering as well low dimensional systems as the mean-field limit and which can serve as platform for other models.

The notion of percolation transition (PT) (*1*) is widely applied in a variety of disciplines as it explains the formation of a spanning cluster connecting two opposite sides of a system in Euclidean space, such as occurs in metal-insulator or sol-gel transitions. Alternatively, percolation can also be interpreted as the formation of a macroscopic cluster in the system, and this concept has been used to model epidemic spreading (*2*) and to community formation within social networks (*3*). These two pictures may be regarded as the same; but, they can lead to different evolution processes in Euclidean space. In this paper, we study an abrupt PT (*4*) from these two perspectives.

One of the models in the more general second category is the classic Erdös and Rényi (ER) (*5*) model, in which the evolution proceeds as follows: Starting with $N$ isolated nodes, an edge is connected between a randomly selected unconnected pair of nodes at each time step. Then, as the number of connected edges is increased, a macroscopic cluster is generated at the percolation threshold, and its size is increased continuously. Recently, the ER model was modified by imposing additionally a so-called product rule or sum rule, which suppresses the formation of a large cluster (*4*). Because of this suppressive bias, the percolation threshold is delayed, and thus, when the giant cluster eventually emerges, it does so explosively. Thus, this model has been called the explosive percolation (EP) model. This result has attracted much interest (*6-23*), including openings towards other subjects, for example, synchronization phenomena (*23*), jamming in the Internet (*24*) and also analysis of real-world networks (25). Initially, this explosive PT was regarded as a discontinuous transition; however, it was recently found that the transition is continuous in the thermodynamic limit (*9*), followed by a mathematical proof (*10*) and extensive supporting simulations (*11-13*). The random graph, in fact represents the mean-field description of the model on a Euclidean lattice. EP problems in Euclidean space have also been considered, and the numerical results suggest discontinuous percolations (*14*). Because of the absence of analytic results, the order of explosive PT in Euclidean space is still not determined yet. Under this circumstance, it is interesting to clarify the order of the explosive PT in Euclidean space and on random graphs in a unified manner.

The product rule or sum rule was inspired by an idea, first mathematically developed in (*26,27*) on the power of multiple choices in random processes. These ideas were refined to what is called an Achlioptas process, in which one chooses the best among randomly given multiple options to avoid the formation of a certain target pattern (*28*). Here, we choose the spanning cluster rather than the giant cluster as target pattern of the PT in Euclidean space. Surprisingly, this choice, while keeping the strategy of choosing the best among several options, enables us to determine analytically the order of the explosive PT for the spanning cluster in the thermodynamic limit. In this spanning-cluster-avoiding (SCA) model, the transition can be either discontinuous or continuous below the upper critical dimension, depending on the number of potential bonds $m$ introduced in the dynamic rule, and it is continuous above the upper critical dimension, i.e., in the mean-field limit. Thus, the appearance of an abrupt PT can be clarified within a unified scheme. Moreover, the analytic results and the methodology used in the SCA model can be a platform to understand the PTs for other models showing abrupt PTs such as the product rule (*4*) and the Gaussian model (*18*). In fact, the general mechanism underneath the abrupt PT is that by throttling spanning, the finite clusters can become very dense so that when they finally merge to a percolating configuration, a substantial fraction of sites is immediately involved in the largest cluster.

We begin by introducing an adequate suppressing rule in Euclidean space. Starting with a $d$-dimensional regular square lattice of linear size $L$ having $N = L^d$ nodes and $N_b = zN$ unoccupied bonds, where $z$ is coordination number divided by two, we randomly choose at each time step $m$ unoccupied bonds. They are classified into two types: bridge and non-bridge bonds. Bridge bonds are those that would form a spanning cluster if occupied. We want to avoid bridge bonds being occupied, and thus one of the non-bridge bonds is randomly selected and occupied (Fig. 1A). If the $m$ potential bonds are all bridge bonds, then one of them is selected randomly and occupied (Fig. 1B). Once a spanning cluster is created, no more restrictions are imposed on the occupation of bonds. This procedure continues until all bonds are occupied. The selection rule among multiple options is inspired by the best-of-$m$ model (*19*). We denote the number and the fraction of occupied bonds as $\ell$ and $t = \ell / zN$, respectively. In fact, in analogy to ordinary percolation, $t$ can be interpreted as an occupation probability, and also as the time of our dynamic system. We define $\ell_{cm}$ as the number of occupied bonds when a bridge bond is occupied for the first time. The percolation threshold is $t_{cm} \equiv \ell_{cm} / N_b$, which for $m > 1$ is larger than the percolation threshold $t_c$ of ordinary percolation.

We performed extensive numerical simulations for various system sizes $L$ and parameter values $m$ to see (i) how the order parameter $G_m(t)$, the fraction of sites (nodes) belonging to the spanning cluster, behaves as a function of $t$ and $m$ and (ii) how the percolation threshold $t_{cm}$ averaged over configurations depends on $m$, $L$, and dimension $d$. Then, (iii) extrapolating these results, we determined the order of the PT under the suppressing rule in the thermodynamic limit.

We now briefly show numerical results and then theoretical results, of which derivations are presented in (29). First, for a given $m$, $G_m(t) = 0$ for $t < t_{cm}$. For $t_c < t < t_{cm}$, a spanning cluster is not created because of the suppressing rule, while it is formed in ordinary percolation. However, for $m > 1$ and $t > t_{cm}$, once one bridge bond is occupied, no further bonds are suppressed and thus one has no effect on the size of the spanning cluster. Thus, $G_m(t)$ follows the curve of $G_1(t)$ for $t > t_{cm}$ (Fig. 2A). Thus, there exists a finite discontinuity $G_1(t_{cm})$ at $t_{cm}$. In the reverse evolution, dynamics proceeds under the bias of sustaining the spanning cluster. For this case, evolution can be understood as opposite procedure from occupation to deletion of bonds. Then, the spanning cluster can sustain up to $1-t_{cm}$, but its size reduces to $N_{BB}/N_b$, where $N_{BB}$ is the number of the bridge bonds. In the thermodynamic limit the fraction $N_{BB}/N_b$ is zero in the interval $[1-t_{cm}, 1-t_c]$ for $d < 6$ (20). Thus, the order parameter behaves as shown in Fig. 2B.

Next, we plot the percolation threshold for systems with linear size $L$, $t_{cm}(L)$ versus $L$ for several $m$ values, and find that $t_{cm}(L)$ decreases (increases) and converges to $t_c$ ($t_{cm}(L \to \infty) \to 1$) as $L$ increases for $m = 2$ (for $m \geq 3$) in two dimensions. More generally, we find that there exists a critical value $m_c(d) = d/(d-d_{BB})$ for $d > d_{BB}$, where $d_{BB}$ is the fractal dimension of the set of bridge bonds (16), such that if $m < m_c$ ($m > m_c$), $t_{cm}(L)$ decreases (increases) and converges to $t_c$ ($t_{cm}(L \to \infty) \to 1$) as $L$ increases. This is shown in Fig.2C for 2-dim, and Figs. S2A and S2C for 3-dim and 4-dim, respectively. The analytic solution for $m_c(d)$ is presented in (29). It is estimated that $m_c(d) \approx 2.55 \pm 0.01$ $(d=2)$, $5.98 \pm 0.07$ $(d=3)$, $16.99 \pm 5.23$ $(d=4)$, $50$ $(d=5)$, and $\infty$ $(d=6)$. For $d = 5$, the standard deviation is larger than the mean value. $d_{BB}$ is equal to $d$ in $d=6$, which is thus the upper critical dimension and the formula for $m_c$ is valid for $d < d_c = 6$. To simulate for non-integer $m$ cases, once we select a unoccupied bond randomly, and if that bond is a bridge bond, then it is occupied with the probability $q(t)^{1-(1/m)}$, where $q(t)$ is the probability that $m$ potential bonds are all bridge bonds, given in (16). Otherwise, it is always occupied. This dynamic process can be implemented without choosing $m$ different unoccupied bonds, but by choosing just one bond.

Subsequently, we check the convergence rates for $t_{cm}(L) - t_c$ and $1 - t_{cm}(L)$ as a function of $L$. We obtain power-law behaviors, $t_{cm}(N) - t_c \sim N^{-1/\bar{\nu}_<}$ and $1 - t_{cm}(N) \sim N^{-1/\bar{\nu}_>}$, where the exponents $\bar{\nu}_<$ and $\bar{\nu}_>$ are derived analytically (*29*) as $1/\bar{\nu}_< = (1 - m/m_c)/(m\zeta + 1)$, and $1/\bar{\nu}_> = (m/m_c - 1)/(m - 1)$. The exponent $\bar{\nu}$ is rewritten as $d\nu$, where $d$ is the spatial dimension, and $\nu$ the exponent characterizing the scaling relation between length scale $L$ and the occupation probability $t$. These results are shown in Fig. 2D for 2-dim and in Figs. S2B and S2D for 3-dim and 4-dim, respectively. We show in (*29*) that the standard deviation for the statistical fluctuations of the critical point $t_{cm}(N)$ behaves in the same manner as $\sigma_< \sim N^{-1/\bar{\nu}_<}$ for $m < m_c$ and $\sigma_> \sim N^{-1/\bar{\nu}_>}$ for $m > m_c$, which is confirmed numerically in Fig.S1B. We remark that at a tricritical point $m_c$, $t_{cm_c}(L \to \infty)$ is finite e.g., $t_{cm_c}(\infty) \approx 0.72$ in two dimensions, which is neither $t_c$ nor unity, and the fluctuation is large and independent of $N$.

Based on the above results, we come to the conclusion that for $d < d_c = 6$, the percolation threshold in the limit $N \to \infty$ is $t_c$ for $m < m_c$, finite $t_{cm}$ at $m = m_c$, and one for $m > m_c$ as shown in Eqs.(8-10) in (*29*). For $d \geq d_c$, $m_c \to \infty$ and for finite $m$, $t_{cm} \to t_c$ (*29*),. We conclude that when $m$ is finite, the PT is continuous in the limit $N \to \infty$. In statistical physics, it is known that mean-field results above the upper critical dimension are equivalent to the solution on sparse random graphs. In this perspective, our result for $d > d_c$ is comparable to previous results for the EP model (*10*) on random graphs.

For the SCA model in the regime $m > m_c$ at $t_{cm}^-(L)$, we find that there are only a few clusters and that they are compact (Fig.3A). Thus, the cluster size distribution at $t_{cm}^-(L)$ decays rapidly in the small-cluster-size region and exhibits a peak in the large-cluster-size region (Fig.3B). The interface between clusters forms naturally along the bridge bonds, and is self-affine. Because of the presence of already macroscopically grown but not yet spanning clusters, the order parameter is increased drastically when occupying a bridge bond. Finally, we remark that for $d \geq d_c$, a discontinuous PT can take place if $m$ varies with the system size $N$. We obtain a characteristic value $m_c \sim \ln N$ such that when $m$ increases with $N$ slower (faster) than $m_c$, the PT is continuous (discontinuous), which occurs at $t_c$ ($t_{cm}(\infty) < 1$). The details are presented in (*29*).

For the product rule (*4*), the nature of the PT is similar to the mean-field behavior of the SCA model in low dimensions such as $d = 2$. Under the best-of-$m$ strategy, when $m$ varies with the system size as $m > m_c \sim \ln N$, clusters are also compact and the number of clusters is limited to a finite value, and thus a discontinuous PT can take place. However, for a fixed $m$ and in the thermodynamic limit, the PT is continuous. The details are presented in (*29*).

The authors thank the anonymous referees for useful comments. This work was supported by an NRF grant awarded through the Acceleration Research Program (Grant No. 2010-0015066) and the Seoul Science Foundation and the Global Frontier Program (YSC).

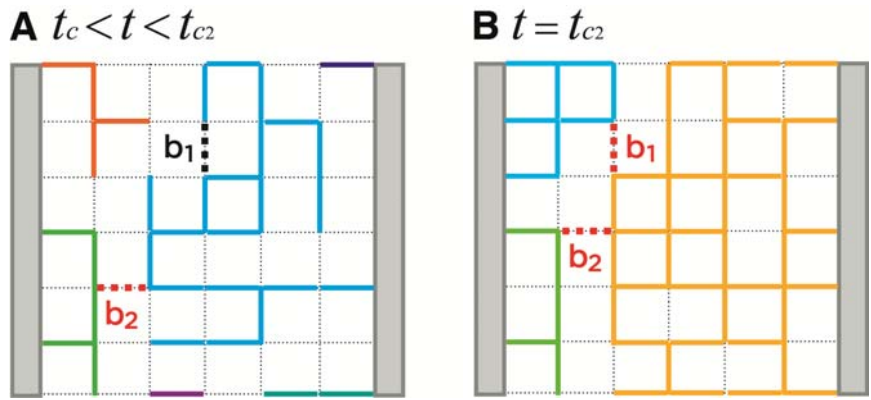

**Fig. 1**. **A**. Dynamics of the SCA model on a square lattice. For the case $m = 2$, two empty bonds, $b_1$ and $b_2$ (shown as dashed lines), are randomly selected. If one of them is a bridge bond ($b_2$), by which a spanning cluster would be formed, then the non-bridge bond ($b_1$) is chosen. **B**. At $t = t_{c2}$, two bridge bonds can be selected for the first time. Then, one of them is taken randomly, and a spanning cluster is formed

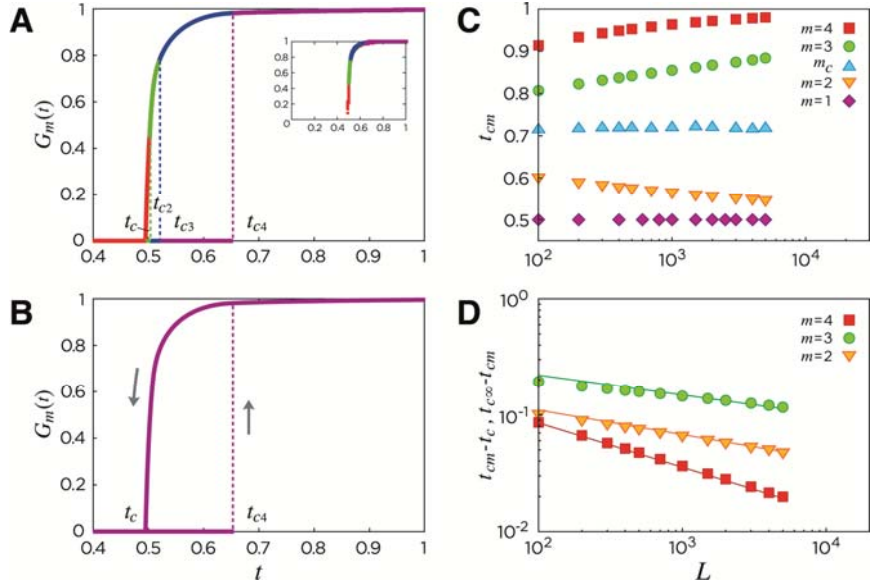

**Fig. 2. A**. Schematic plot of the spanning cluster size $G_m(t)$ versus $t$ the number of attached bonds per $N_b$ for the SCA model with $m=1$ (red), 2 (green), 3 (blue), and 4 (purple). Inset: Same plot with real data, which are obtained after averaging over the samples containing non-zero $G_m(t)$ at each time $t$. Data are obtained for a system size of $N=10^6$ in two dimensions. **B**. Hysteresis curve of the order parameter in forward and backward evolution. **C**. Plot of $t_{cm}(L)$ versus $L$ for various $m$ values. **D**. Plot of $t_{cm}(L) - t_c$ for $m=2 < m_c$ and $1 - t_{cm}(L)$ for $m=3$ and $4 > m_c$ vs. $L$. Solid lines are guidelines with the slopes theoretically predicted. All data are averaged over $10^4$ configurations.

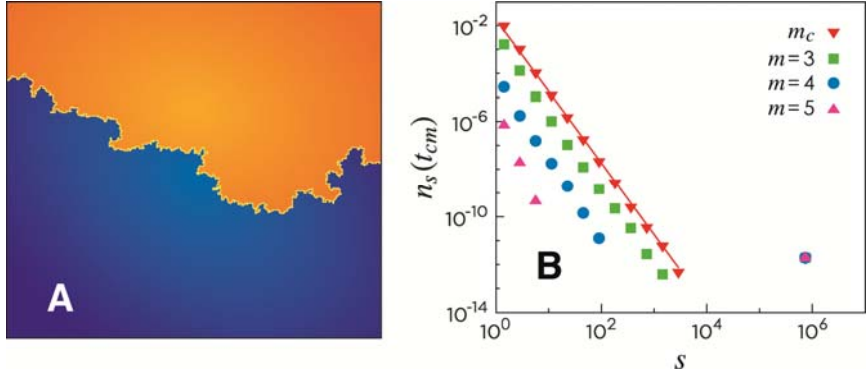

**Fig. 3. A.** Plot of the giant and the second largest clusters just before the percolation threshold $t^-_{c4}$ for $m = 4$. Clusters are compact and the boundary is self-affine with the fractal dimension $d_{BB} \approx 1.22$. In general, for sufficiently large $m$ only very few clusters remain at $t^-_{cm}(N)$. **B.** Plot of $n_s$ versus $s$ at $t^-_{cm}$ for $m = 3$, 4, 5, and $m_c \approx 2.55$ in two dimensions. Simulations are performed for $N = 10^6$, and the data are averaged over $10^3$ configurations. Solid line is a guideline with slope $-3$.

## Supplementary Materials

**Percolation transitions for $d < d_c$: Analytic solutions for $t_{cm}$**

We consider the probability $q(\ell)$ that all randomly chosen $m$ unoccupied bonds are bridge bonds at $t$, given by $\ell = tN_b$. This probability is obtained as

$$q(\ell) = \left[\frac{N_{BB}}{N_b(1-t)}\right]^m \sim N^{-m/m_c}\left[\frac{(t-t_c)^\zeta}{(1-t)}\right]^m, \quad (1)$$

where $q(\ell) = 0$ for $\ell \leq t_c N_b$, and $N_{BB}$ is the number of bridge bonds, given by

$$N_{BB} \sim \begin{cases} L^{1/\nu_0} & \text{at } t = t_c, \\ L^{d_{BB}}(t-t_c)^\zeta & \text{at } t > t_c. \end{cases} \quad (2)$$

The exponent $\nu_0$ is the correlation length exponent for ordinary percolation transition (PT), $d_{BB}$ is the fractal dimension of a pattern formed by bridge bonds (20), and the exponent $\zeta$ is an exponent describing the transition behavior of $N_{BB}$. As $N_{BB}$ increases, bridge bonds form a chain or a surface in two and three dimensions, respectively, and thus the fractal dimension of bridge bonds is bounded as $d - 1 < d_{BB} < d$. $m_c$ is denoted as

$$m_c(d) = \frac{d}{d - d_{BB}}, \quad (3)$$

across which the order of the PT changes. Numerical values of $m_c(d)$ for various dimensions are listed in Table S1.

Next, we calculate the probability that the system reaches a percolating state at $\ell$ for the first time, which is given by

$$Q(\ell) = \left[\prod_{k=0}^{\ell-1}(1-q(k))\right] q(\ell) \approx e^{-\sum_{k=0}^{\ell-1} q(k)} q(\ell). \tag{4}$$

We call $\ell_{cm}$ the value at which $Q(\ell)$ has its maximum. Then, the relation

$$q(t) = \frac{1}{N_b} \frac{\partial \ln q(t)}{\partial t} \tag{5}$$

holds at $t_{cm} = \ell_{cm}/N_b$. $t_{cm}$ depends on the system size $N$. We find that $t_{cm}(N)$ behaves differently for the cases $m < m_c$ and $m > m_c$. When $1 < m < m_c$, $t_{cm}(N) - t_c$ decreases as a power law $\sim N^{-1/\bar{\nu}_<}$, and when $m > m_c$, $1 - t_{cm}(N)$ decreases as $\sim N^{-1/\bar{\nu}_>}$. The exponents are obtained by inserting Eq. (1) into Eq. (5) as

$$\frac{1}{\bar{\nu}_<} = \frac{1}{m\zeta + 1}\left(1 - \frac{m}{m_c}\right), \tag{6}$$

and

$$\frac{1}{\bar{\nu}_>} = \frac{1}{m-1}\left(\frac{m}{m_c} - 1\right). \tag{7}$$

We remark that at $m_c$, $q(\ell) \sim N^{-1}\left[(t-t_c)^\zeta/(1-t)\right]^{m_c}$, which may be written as $q(\ell) = N^{-1}\tilde{q}(\ell)$, where $\tilde{q}(t)$ is independent of $N$. Then, $\tilde{q}(t)$ satisfies the equation $\tilde{q}(t) = \partial \ln \tilde{q}(t)/\partial t$ and its solution is $t_{cm_c}$. Thus, $t_{cm_c}$ is finite in the thermodynamic limit, which is neither $t_c$ nor unity. Moreover, $q(\ell) \sim O(1/N)$ at $t_{cm_c}$, and thus the probability to actually occupy a bridge bond is non-negligible.

**Percolation transitions for $d \geq d_c$**

When $d = d_{BB}$ for $d \geq d_c$, the probability $q(t)$ in Eq. (1) is independent of the system size $N$ as

$$q(t) \sim \left[\frac{(t-t_c)^\zeta}{(1-t)}\right]^m \equiv f^m(t). \tag{8}$$

Then, for finite $m$, following a similar step as used to derive Eq. (6) from Eq. (5), we obtain that

$$t_{cm}(N) - t_c \sim N^{-1/(m\zeta+1)}. \tag{9}$$

Then a continuous PT occurs at $t_c$ in the thermodynamic limit.

If $m$ varies with $N$ as $m \sim \ln N$, then $t_{cm} - t_c$ is independent of $N$. That is, $t_{cm}$ is finite. Then, a discontinuous PT occurs at finite $t_{cm}$, which is neither $t_c$ nor unity. Furthermore, when $m$ is increased faster than $\ln N$ as $N$ is increased, $t_{cm}$ increases. In this case, we rewrite Eq. (5) as

$$\ln f = -\frac{\ln N}{m+1} + \frac{\ln f'}{m+1} + \text{const}, \qquad (10)$$

at $t = t_{cm}$, where the prime denotes the first derivative with respect to $t$. For the case $\lim_{N \to \infty} \ln N / (m+1) \to 0$, we obtain a finite $t_{cm}$ which satisfies the relation $(t_{cm} - t_c)^\zeta = C(1 - t_{cm})$, where $C$ is const. Actually, this $t_{cm}$ is the percolation threshold of the bridge percolation transition (20).

**Statistical fluctuations of the percolation threshold**

Next, we consider the statistical fluctuations of the critical point $t_{cm}$. To quantify the fluctuations, we consider the percolating probability distribution function $Q(t)$ around the critical point $t_{cm}$. The standard deviation $\sigma(m, N)$ of $t_{cm}$ is calculated using the relation

$$1/\sigma^2 \sim (\ln Q(t))'' \qquad (11)$$

at $t_{cm}$, where the double prime denotes the second derivative with respect to $t$. Then, we obtain that

$$\sigma \sim \begin{cases} N^{-1/\bar{\nu}_<} & \text{for } m < m_c, \\ N^{-1/\bar{\nu}_>} & \text{for } m > m_c. \end{cases} \qquad (12)$$

Thus, for $m < m_c$ ($m > m_c$), the fluctuations are relatively large near $m_c$ and become smaller with decreasing $m$ (increasing $m$) for a fixed system size. The standard deviations shrink with increasing $N$ for a given $m$. However, when $m = m_c$, the standard deviation remains constant, independent of the system size. Thus, it generically appears that the explosive percolation threshold fluctuates heavily from sample to sample at the tricritical point $m_c$.

The theoretical prediction is confirmed by simulation data, shown in Fig. S1. On the other hand, from Eq. (12), we can find that the exponent $\nu$ characterizes the scaling relation between length $L$ and the occupation probability $t$.

**Percolation transitions in the product rule model**

Here we study the explosive PT for the product rule (PR) model under the best-of-$m$ strategy in Euclidean space, in which the dynamics proceeds under the suppression against the formation of large cluster. At each time, $m$ distinct potential bonds are selected randomly, and a set of the products of two clusters connected by each of those potential bonds are calculated. If a selected potential bond is an intra-bond, then the product is taken as the square of that cluster size. The potential bond producing the minimum of those products is actually added, and the other $m-1$ potential bonds are discarded. Previous studies focused on the case $m=2$ in Euclidean space (6, 13, 14). Here, we want to see the behavior of PT for general $m$, and compare it with the results for the SCA model.

First, in Fig. S3A, for a given system size, for example, $L=200$ in two dimensions, we examine the behavior of the giant cluster size per node $G_m(t)$ as a function of time $t$ for various $m$. The transition point $t_{cm}(L)$ is delayed as $m$ is increased. Interestingly, the increasing pattern of $G_m(t)$ for relatively small $m$ such as $m=2-8$ is different from that for relatively large $m$ such as $m \geq 20$: (i) For those small $m$, $G_m(t)$ increase sharply. However, the jump size decreases with increasing the system size (Fig. S3E). For this case, large clusters are not compact, but contain isolated small clusters within them at the onset of the abrupt change of the order parameter, denoted as $t_{cm}^-$ (Fig. S4A). The cluster size distribution at $t_{cm}^-$, decays in a power-law manner in small-cluster-size region, but shows a bump in large-cluster-size region (Fig. S4B). (ii) For those large $m$, $G_m(t)$ increases irregularly, caused by the average of plateaus from different configurations (Fig. S3B). Plateaus are formed when intra-cluster bonds are added to the system. During the transient period, small clusters merge to surrounding large clusters, and these merging dynamics leads to the formation of compact clusters at $t_{cm}^-$ as shown in Fig. S4C. The interface between two compact clusters is self-affine with the same fractal dimension as that of bridge bonds in the SCA model (Fig. S5). Cluster sizes of such compact clusters are almost homogeneous as shown in the cluster size distribution in Fig. S4D, because smaller (larger) clusters are more likely to merge to others (be excluded from merging). We find that the jump size of the order parameter remains finite as the system size is increased (Fig. S3F).

The number of distinct clusters $N_{cl}(t_{cm}^-)$ at $t_{cm}^-$ exhibits distinct feature. (i) When $m$ is small such as $m = 3$, the extensive relationship $N_{cl}(t_{cm}^-) \sim N$ holds, whereas (ii) when $m$ is large, a crossover behavior occurs: For small $N$, the sub-extensive or the system-size independent relationship $N_{cl}(t_{cm}^-) \sim N^\alpha$ with $\alpha < 1$ or $\alpha \approx 0$ appears, whereas for large $N$, the extensive relationship $N_{cl}(t_{cm}^-) \sim N$ does. Such behaviors in 2-dim are shown in Fig. S6A. The crossover point, denoted as $N^*$, is scaled as $N^* \sim e^{am}$, where $a$ is const. When $N$ is fixed, the crossover occurs at $m^* \sim \ln N$. Thus, as $N \to \infty$, $m^* \to \infty$. The phase diagram is shown in Fig. S6B. This result is intrinsic, independent of dimension. We argue that when the number of clusters linearly depends on (is sub-extensive of) system size, the jump size of the order parameter decreases (finite). Thus, for a finite fixed $m$, the PT is continuous in the limit $N \to \infty$ even in low dimensional systems for the PR model. However, if $m$ varies with the system size with the constraint $m \geq \ln N$, then the percolation transition could be discontinuous. This result is the same as the mean-field result above the upper critical dimension $d_u = 6$ for the SCA model. The crossover behavior is also observed on random graph (Fig. S6C and D).

Here, we present hand-waving argument to support the above numerical results. We consider the case (i) with relatively small $m$. We show snapshots of the system at several time steps to display how the system evolves, finding that small clusters still remain above the percolation threshold but large clusters merge, and make eventually a largest cluster (Fig. S7). As $m$ is increased, the bump size in the large-cluster-size region is increased, whereas the small-cluster-size region is reduced (Fig. S8). We denote the characteristic cluster size as $s^*$, which divides those two regions. We calculate the probability $p_s$ ($p_l$) that a cluster with size $s < s^*$ ($s > s^*$) is selected by random selection as

$$p_s(t) = \frac{\int_1^{s^*} ds \, n_s s f_s}{(1-t)} \quad \text{and} \quad p_l(t) = \frac{\int_{s^*}^{\infty} ds \, n_s s f_l}{(1-t)}, \tag{13}$$

where $n_s(t)$ is the cluster size distribution at $t$, $f_s$ and $f_l$ are the mean densities of unoccupied bonds in small and large clusters, respectively. $(1-t)$ is the density of unoccupied bonds in the system. The relation $p_s + p_l = 1$ holds. Then, the probability that all $m$ potential bonds are selected in the bump region is given as $P_{l,m}(t) \equiv p_l^m(t)$. If the condition $P_{l,m} < N^{-1}$ is satisfied, then the probability that an unoccupied bond is actually added in a cluster with size $s > s^*$ is negligible. Thus, a bond in small clusters is actually occupied, and as a result, clusters become compact and $N_{cl}(t_{cm}^-) \sim N^\alpha$ with $\alpha < 1$ at $t_{cm}^-$. This condition may be written as $m > m_c \equiv \ln N / |\ln p_l|$ unless $p_l = 0$. For $m < m_c$, small clusters co-exist, and then $N_{cl}(t_{cm}^-) \sim N$ at $t_{cm}^-$. We note that such different cluster shapes for $m < m_c$ and $m > m_c$ can be seen in the SCA model in Fig. S9.

When $m > m_c$, clusters are compact, and they merge as time passes. We consider the cluster merging dynamics (Fig. S9) starting from the time $t_{cm}^-$ to $t_{cm}^+$, at which single macroscopic-scale cluster is left. Because the cluster size distribution is in narrow bell shape (Fig.S4D), we assume that all cluster sizes are uniform for simplicity. Then, the linear size of each cluster is $L_{cl}(t) \sim (N/N_{cl}(t))^{1/d}$ and the total length of the interfaces is scaled as $M_{BB}(t) \sim N_{cl}(t) L_{cl}^{d_{BB}}(t)$. Cluster merging process takes place when an inter-cluster bond is actually occupied, which occurs with the probability,

$$Q^{(\mathrm{PR})}(t) \sim \sum_{k=1}^{m} \frac{k}{m} \binom{m}{k} \left(\frac{M_{BB}}{N_b(1-t)}\right)^k \left(1 - \frac{M_{BB}}{N_b(1-t)}\right)^{m-k}$$
$$\sim \frac{1}{1-t} \left(\frac{N_{cl}(t)}{N}\right)^{1-\frac{d_{BB}}{d}}. \tag{14}$$

After two clusters are merged, it is assumed that remaining clusters are reorganized and their sizes are again uniform with a bigger size but their number is reduced. This assumption is viewed in ensemble average perspective. Such processes are repeated. The decreasing rate of the number of clusters is given as

$$\frac{1}{N_b} \frac{dN_{cl}}{dt} = -Q^{(\mathrm{PR})}(t). \tag{15}$$

Because the abrupt phase transition occurs in a short time interval, we may approximate that $1/(1-t) \approx 1/(1-t_{cm}^-)$ for $t_{cm}^- < t < t_{cm}^+$. Using Eqs. (14) and (15), and the boundary conditions $N_{cl}(t_{cm}^-) \sim N^\alpha$ and $N_{cl}(t_{cm}^+) = 1$, we obtain that

$$t_{cm}^+ - t_{cm}^- \sim N^{-(1-\alpha)d_{BB}/d}. \tag{16}$$

We have confirmed the exponent numerically in the inset of Fig. S3F. Since the exponent is $\frac{(1-\alpha)d_{BB}}{d} > 0$ for $\alpha < 1$, $t_{cm}^+ - t_{cm}^-$ is reduced to zero as $N \to \infty$, whereas the jump size of the order parameter remains as $O(1)$. Thus, the transition becomes sharper as the system size is increased. However, when the system size exceeds $N^*$, $\alpha = 1$. Thus, the transition becomes continuous. Therefore, the criterion whether $\alpha < 1$ or $\alpha = 1$ in the thermodynamic limit is a good indicator to determine the type of PT.

It is noteworthy that the PT for the PR model is reduced to the mean-field result for the SCA model, which is caused by the following reason. In the SCA model, the cluster merging dynamics is mainly determined by bridge bonds, of which the number is smaller than $O(N)$ below the upper critical dimension, but is in order of $O(N)$ in the mean-field limit. In the PR model, however, the cluster merging dynamics is determined not only by selection of unoccupied bonds on the interface, but also of intra-cluster bonds, and their total number is $O(N)$. Thus, the PT for the PR model is reduced to the mean-field behavior of the SCA model even in low dimensions.

**Percolation transitions in the Gaussian model**

The Gaussian model (*18*) was introduced to search a discontinuous percolation transition, which is defined as follows: At each time step, an empty bond is selected randomly, and is occupied with the probability,

$$p \sim \exp\left[-w\left(\frac{s-\bar{s}}{\bar{s}}\right)^2\right], \tag{17}$$

where $s$ is the cluster size created by occupying the bond and $\bar{s}$ is the average cluster size in the system that would be formed after the empty bond is occupied. $w$ is a control parameter defined in the region $w > 0$. When $w = 0$, the model is reduced to the ordinary percolation model. When $w = 1$, this model displays compact clusters at the onset of the PT, and the transition is discontinuous, independent of spatial dimension (*18*).

Here, we examine the number of clusters $N_{cl}(t_{cm}^-)$ as a function of the system size $N$ at the onset of the PT. It is expected that for a fixed $w$, $N_{cl}(t_{cm}^-)$ increases linearly with $N$ for small system sizes, and sub-extensive for large system sizes. Indeed, we find that such crossover behaviors appear in two dimensions (Fig. S11). Thus, for finite $w$, the percolation transition is discontinuous in the thermodynamic limit for the Gaussian model.

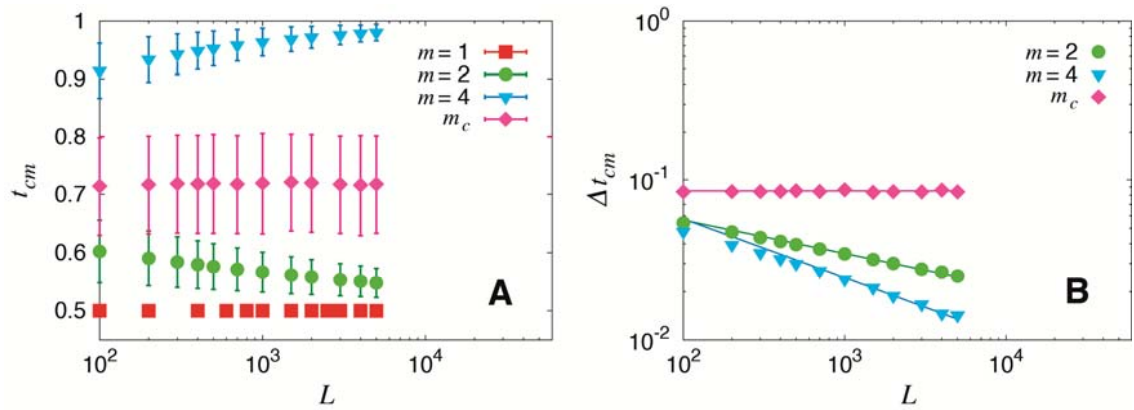

**Figure S1.**
(**A**) Plot of the percolation thresholds with variance versus system size $L$ in two dimensions for the SCA model. (**B**) Plot of the standard deviations of $t_{cm}$ versus system size $L$ in two dimensions. They decay in a power-law manner for $m \neq m_c$, and are flat for $m = m_c$. Solid lines are guidelines with theoretically obtained slopes. All data are obtained after averaging over $10^4$ configurations.

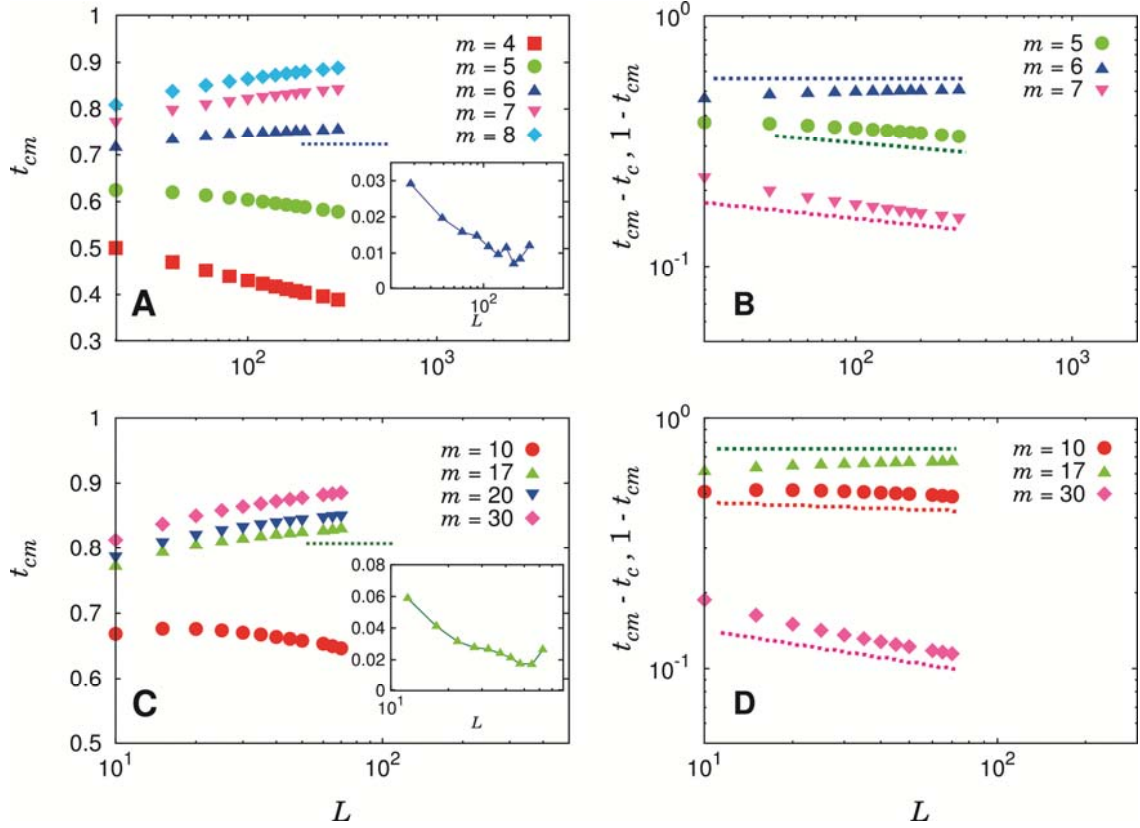

**Figure S2.**
(**A**) and (**C**) Plot of the percolation threshold $t_{cm}$ versus system size $L$ for various numbers of potential bonds $m$ in three (**A**) and four (**C**) dimensions for the SCA model. Dotted lines are straight lines. Inset: Successive slopes of $t_{cm}$ vs. $L$ for $m_c \approx 6$ (**A**) and $m_c \approx 17$ (**C**). They tend to decrease to zero, which implies that $t_{cm}$ converges to a finite value. We also checked that the successive slopes of $1 - t_{cm}$ and $t_{cm} - t_c$ decrease to zero at $m_c$. This indicates that $t_{cm}$ is finite, neither $t_c$ nor 1. (**B**) and (**D**) Plot of scaling relations, $t_{cm} - t_c$ vs. $L$ for $m < m_c$ and $1 - t_{cm}$ vs. $L$ for $m > m_c$ in three (**B**) and four (**D**) dimensions. Dotted lines are guidelines with theoretically obtained slopes. All data are obtained after averaging over $10^4$ configurations.

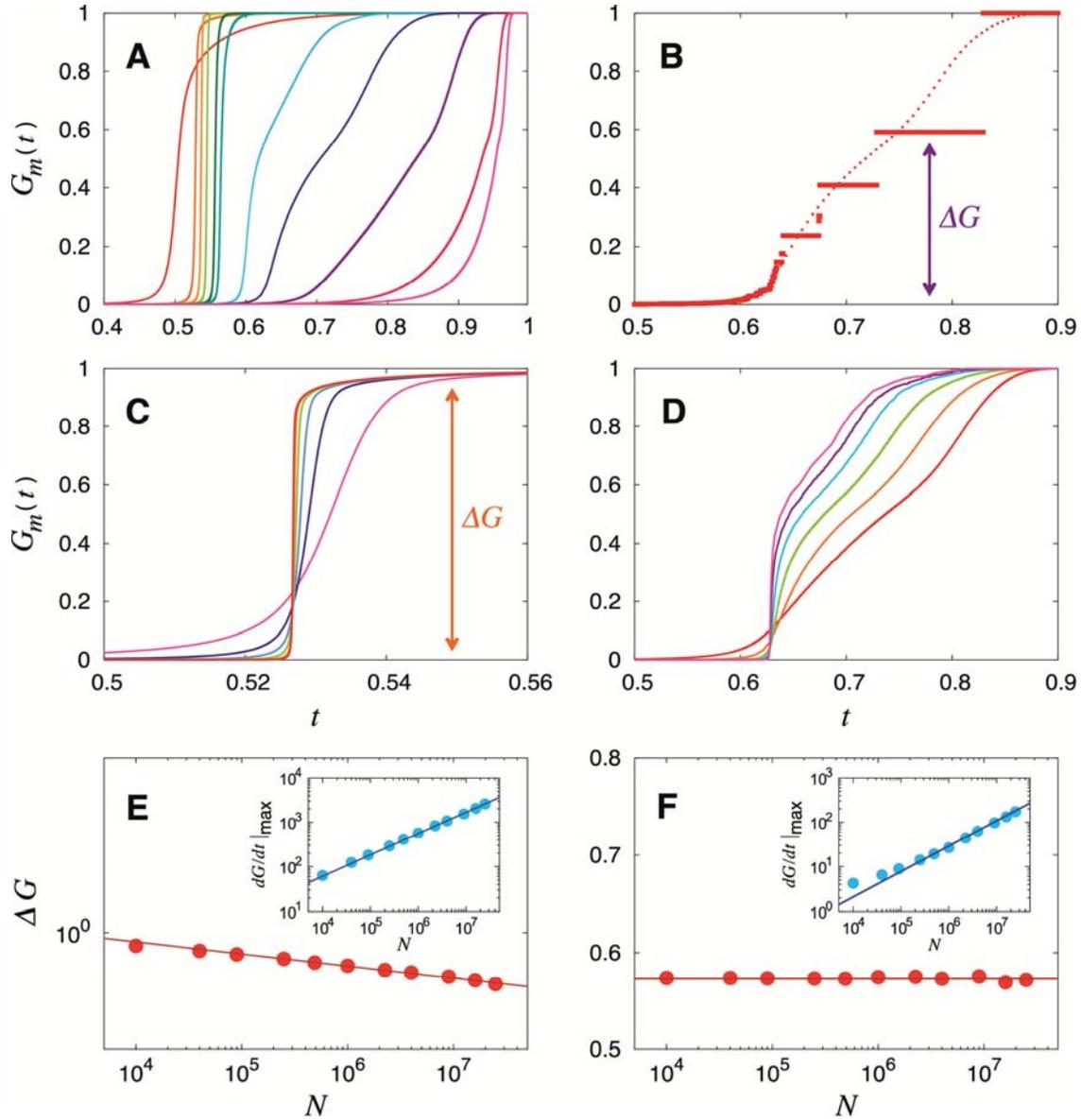

**Figure S3.**
(**A**) Plot of the giant cluster size $G_m(t)$ vs. time $t$ after ensemble average for various $m$ for the PR model in two dimensions. Data are obtained for $m = 1, 2, 3, 5, 8, 10, 20, 30, 50, 100$, and $300$ from the left to the right with a fixed size $L = 300$. (**B**) Similar plot of $G_m(t)$ for $m = 30$ and $L = 200$ after ensemble average (dotted line) and for a single configuration (solid lines). (**C**) and (**D**) Plot of the giant cluster $G_m(t)$ vs. time $t$ for fixed $m = 2$ (**C**) and $m = 30$ (**D**), but various system sizes $L = 100, 300, 700, 1500, 3000$, and $5000$ from the right to the left. As the system size increases, a giant cluster grows more drastically. (**E**) Plot of the jump size vs. system size $N$ for $m = 2$. For $m < m_c$, jump size is measured as the height of $G_m(t)$ just after the abrupt growth. Solid line is a guideline with slope $-0.02$. Inset: Plot of the maximum value of $dG_m(t)/dt$ vs. system size $N$. Solid line is a guideline with slope $0.48$. (**F**) Plot of the jump size vs. system size $N$ for $m = 30$. In this case ($m > m_c$), jump size is measured as the height when a large-scale plateau takes place. The estimated jump size is independent of $N$. Inset: Plot of the maximum value of $dG_m(t)/dt$ vs. system size $N$. Solid line has slope $0.57$, consistent with the theoretical value $(1-\alpha)d_{BB}/d$. Data for $G_m(t)$ and $\Delta G$ are obtained after averaging over $10^4$ configurations.

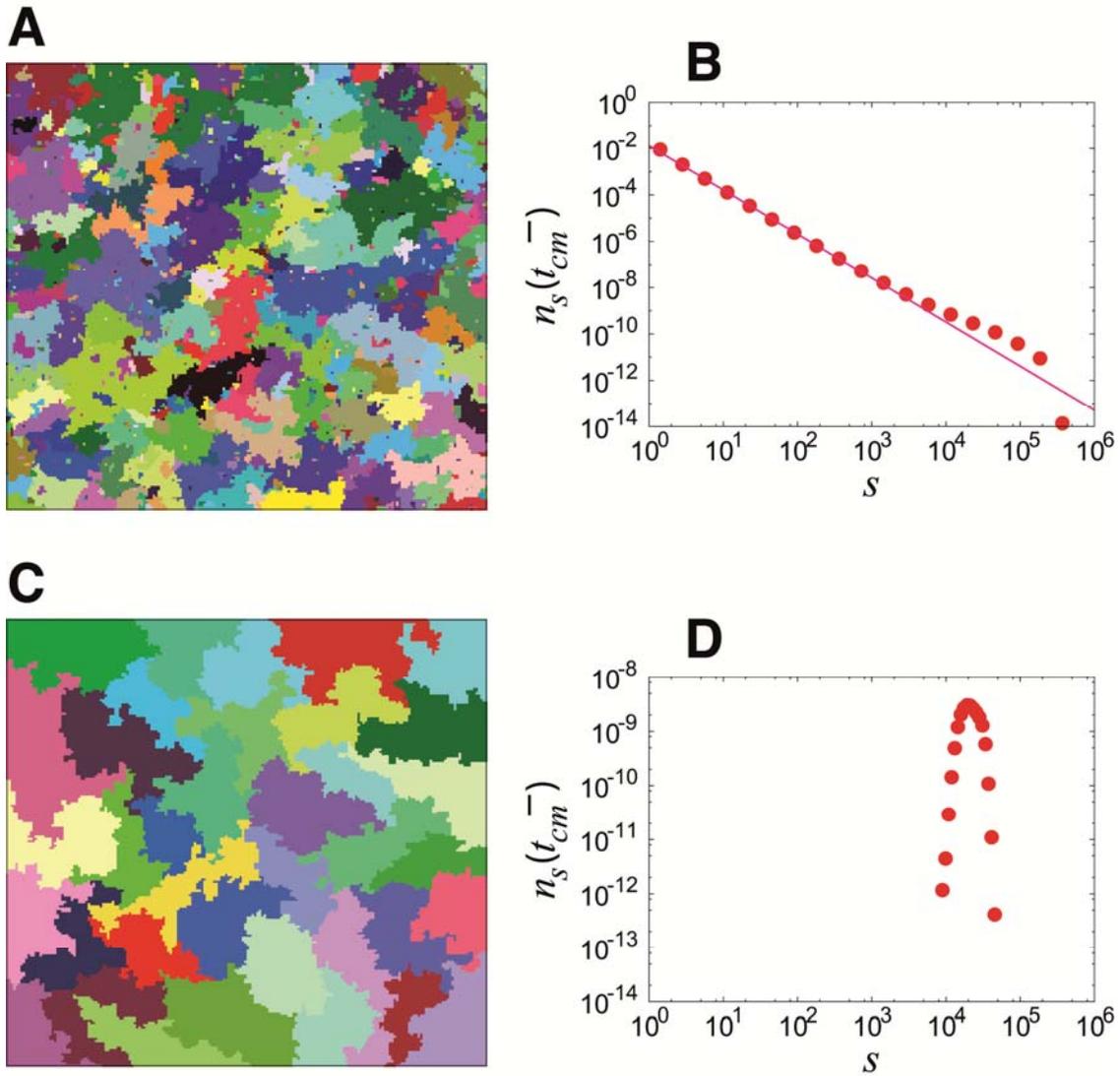

**Figure S4.**

Snapshots of clusters for the PR model at time $t_{cm}^-$ at which the order parameter begins to increase drastically. The data are obtained from two dimensional systems with linear size $L = 200$, but for different $m$ values. For small $m = 2$ (**A**), clusters contain small-sized clusters within them. For large $m = 30$ (**C**), clusters are compact. Plot of the cluster size distributions for $m = 2$ (**B**) and $m = 30$ (**D**) at $t_{cm}^-$. Data in (**B**) and (**D**) are obtained after averaging over $10^3$ configurations in two dimensions with linear size $L = 10^3$. Solid line in (**B**) has slope $-1.95$.

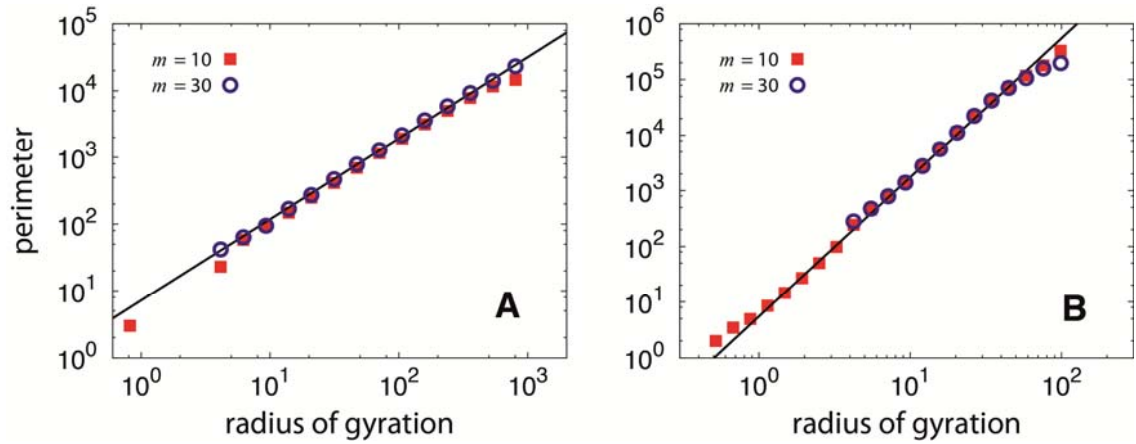

**Figure S5.**

Plot of the number of unoccupied bonds which inter-connect one cluster and its neighbor if occupied (length of perimeter) vs. radius of gyration of each cluster for $m=10$ and 30 in two (**A**) and three (**B**) dimensions at $t_{cm}^{-}$ for the PR model. Thus, the slope means the fractal dimension of perimeter of clusters. Solid lines are guidelines with slopes 1.22 (**A**) and 2.5 (**B**), respectively. Simulation results are obtained after averaging over $10^3$ configurations.

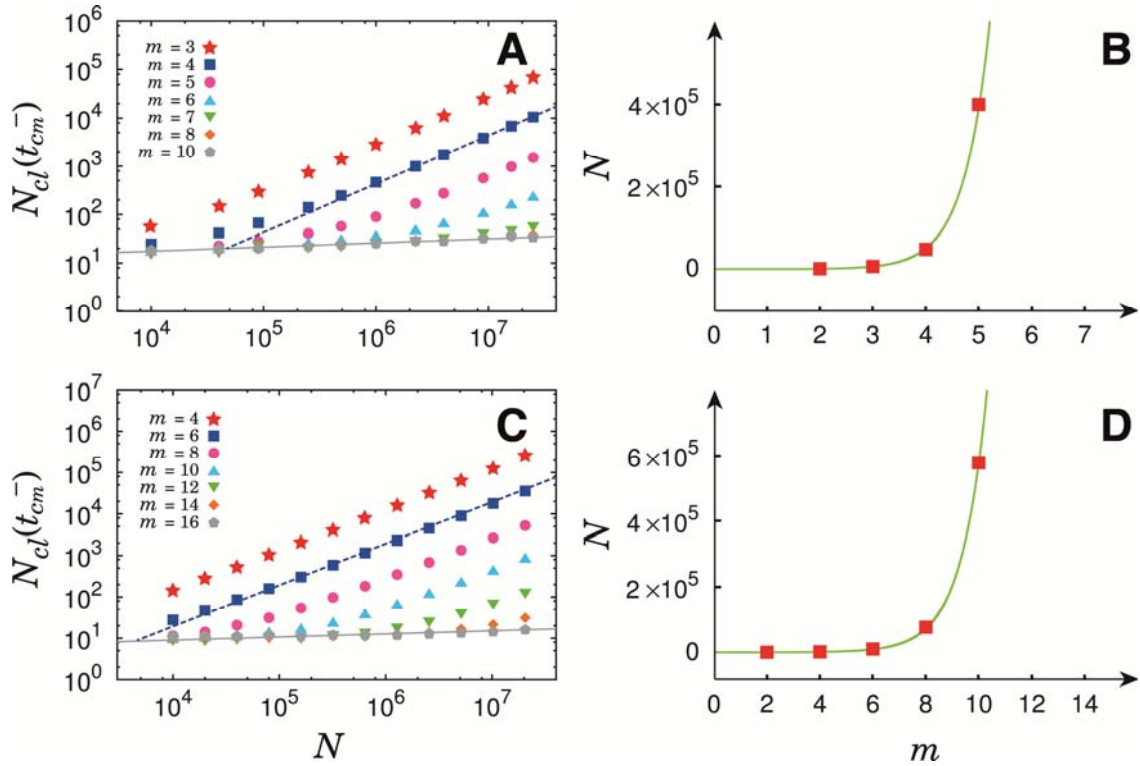

**Figure S6.** (**A**) Plot of the number of clusters at $t = t_{cm}^-$ vs. system size $N$ for several values of $m$ for the PR model in two dimensions (**A**) and random graphs (**C**). Solid (Dashed) lines are guidelines with slopes 0.05 (one). A crossover behavior occurs between the behaviors $N_{cl} \sim N^\alpha$ ($\alpha \approx 0.05$) for $N < N^*$ and $N_{cl} \sim N$ for $N > N^*$, at the point $N^* \sim e^{am}$ ($a = $ const.), which are represented by the solid curves in (**B**) and (**D**). Red squares in (**B**) and (**D**) are estimated values of $N^*$ from the numerical data in (**A**) and (**C**). All data are obtained after averaging over $10^4$ configurations.

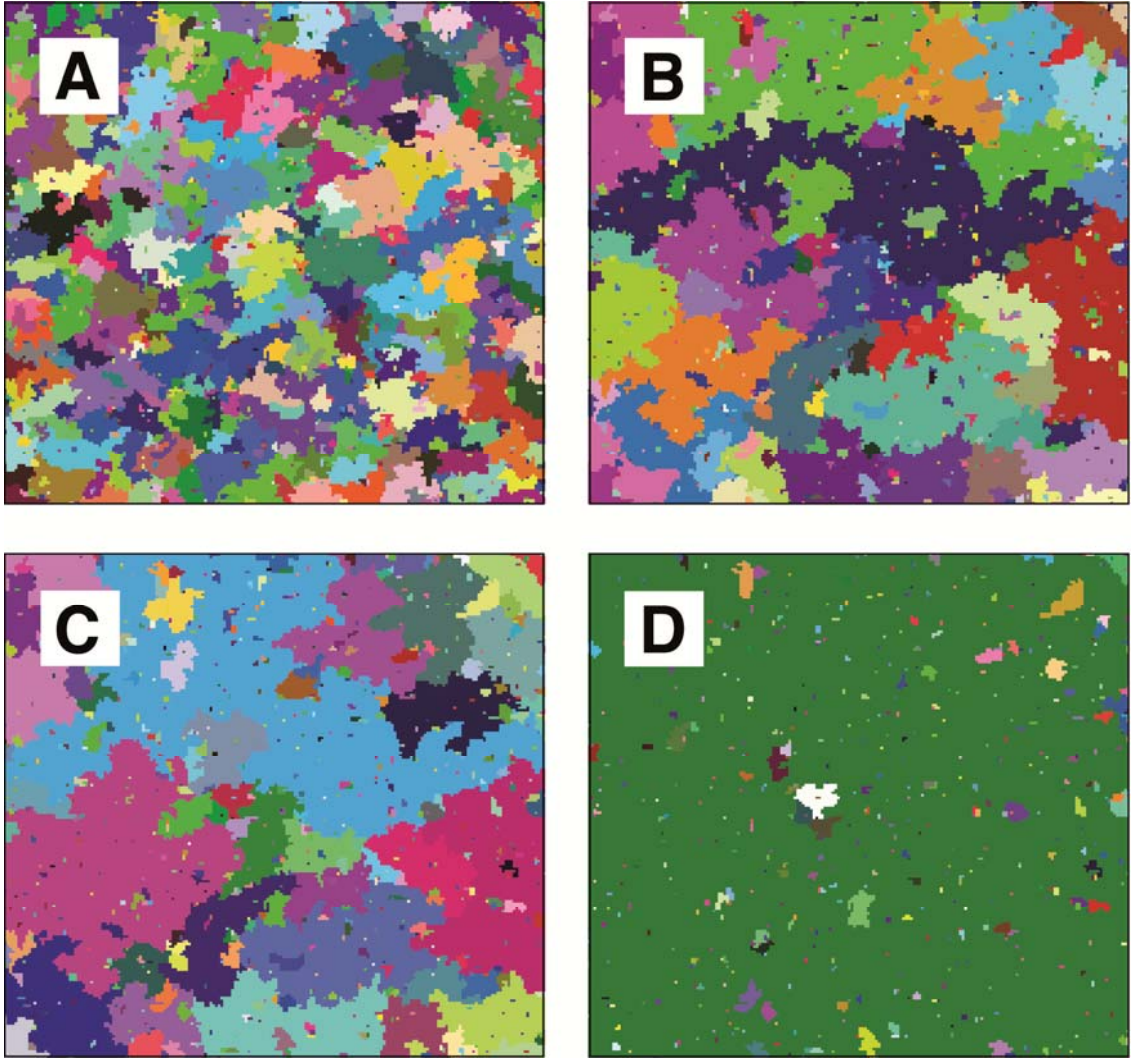

**Figure S7.** Snapshots of clusters in the system with $L = 200$ at several time steps for $m = 2 < m_c$ for the PR model at $t_x$ in ($t_c < t_x < t_{cm}^-$) (**A**), $t_{cm}^-$ (**B**), $t_{cm}$ (**C**), and $t_{cm}^+$ (**D**). Even after the giant cluster is formed, small-sized clusters still remain in the system.

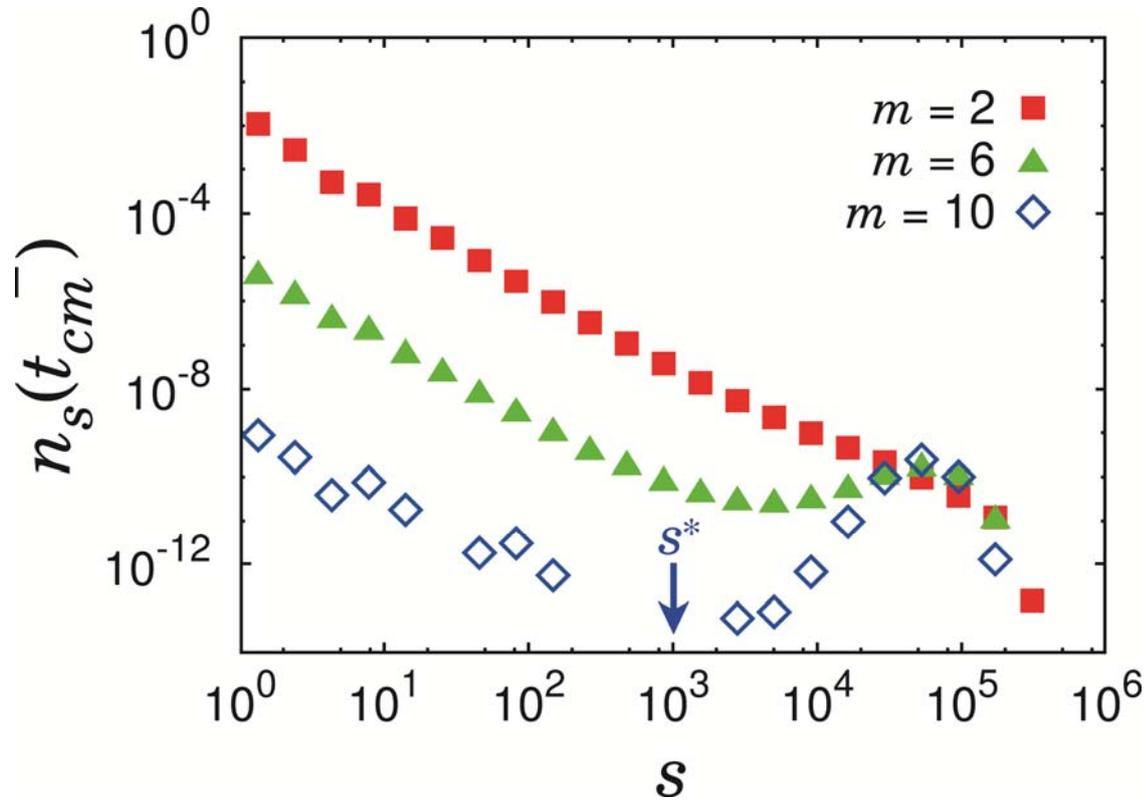

**Figure S8.** Plot of the density of $s$-sized clusters at $t_{cm}^-$ vs. cluster size $s$ in 2-dim with system size $L = 10^3$ for several $m$. As $m$ is increased, bump size is increased, and the small-cluster-size region is reduced. Here $s^*$ is the characteristic size beyond which the bump is distinct. All data points are obtained after averaging over $10^4$ configurations.

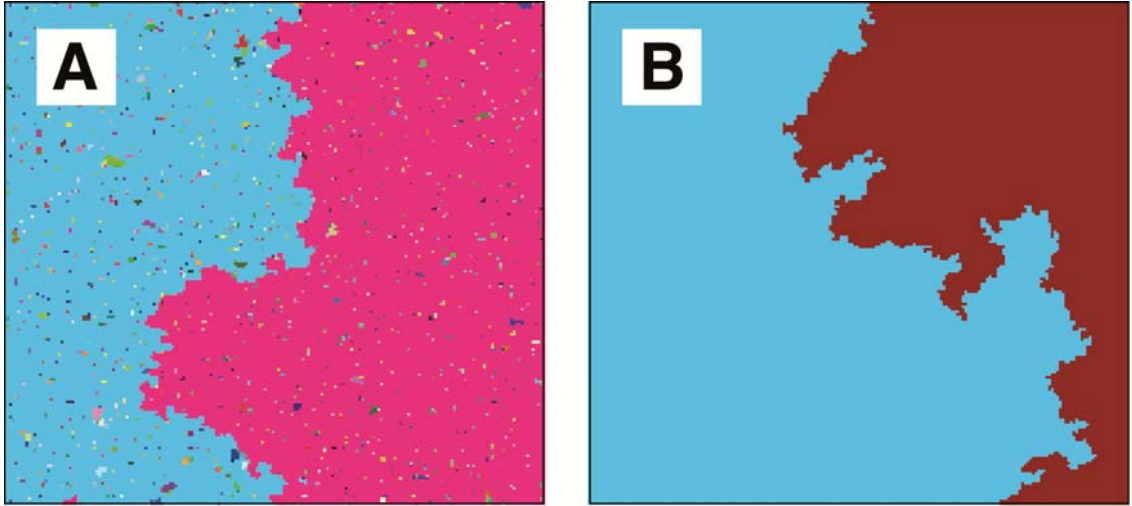

**Figure S9.** Snapshots of clusters for the SCA model when $m = 2$ (**A**) and 5 (**B**) in the system with $L = 200$ at $t^-_{cm}$. Two large clusters are not compact in (**A**) but include small clusters, but compact in (**B**).

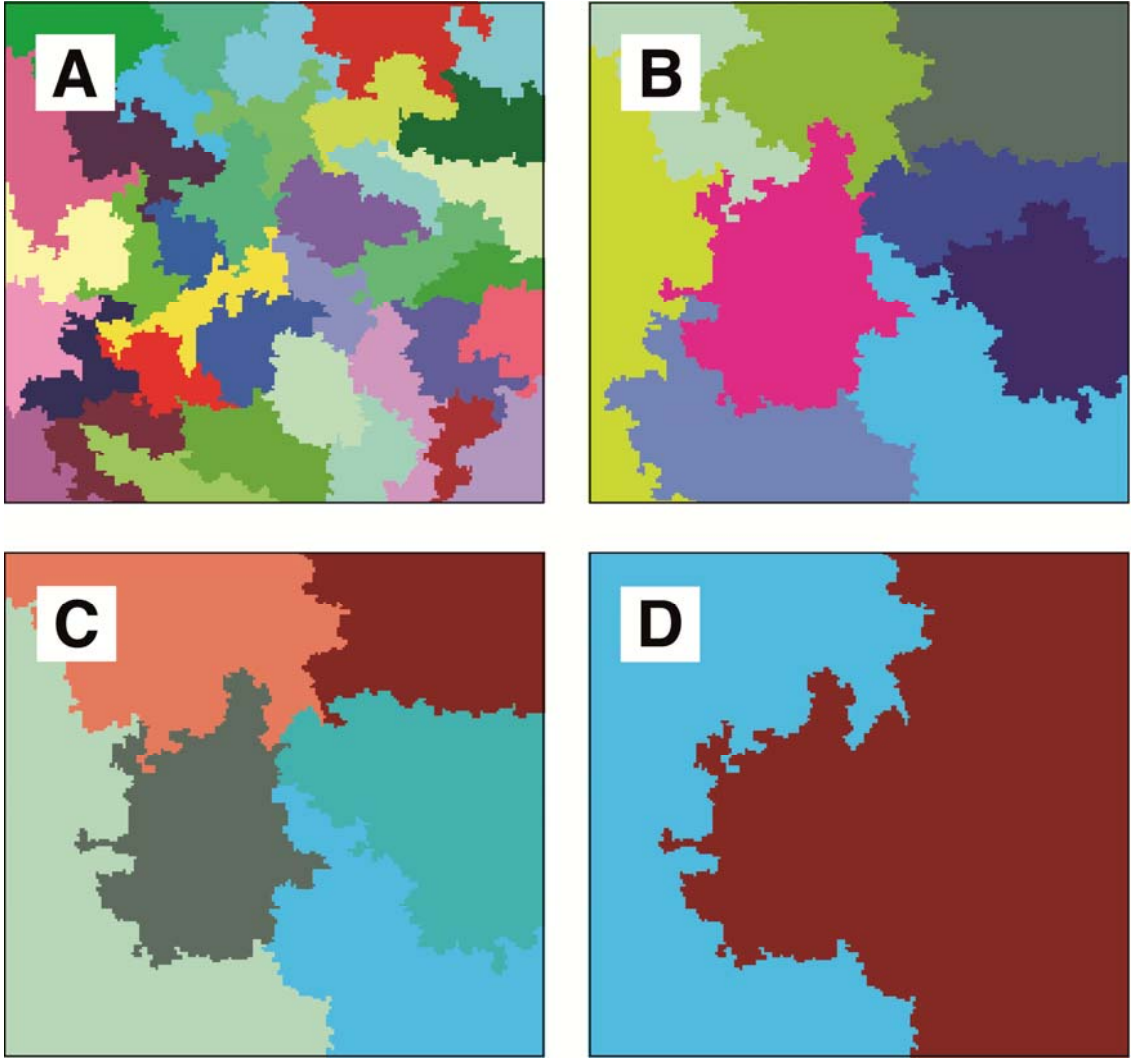

**Figure S10.** Snapshots of clusters for $m = 30 > m_c$ in a system with $L = 200$ for the PR model at several time steps when jumps arise as shown in Fig. S3B.

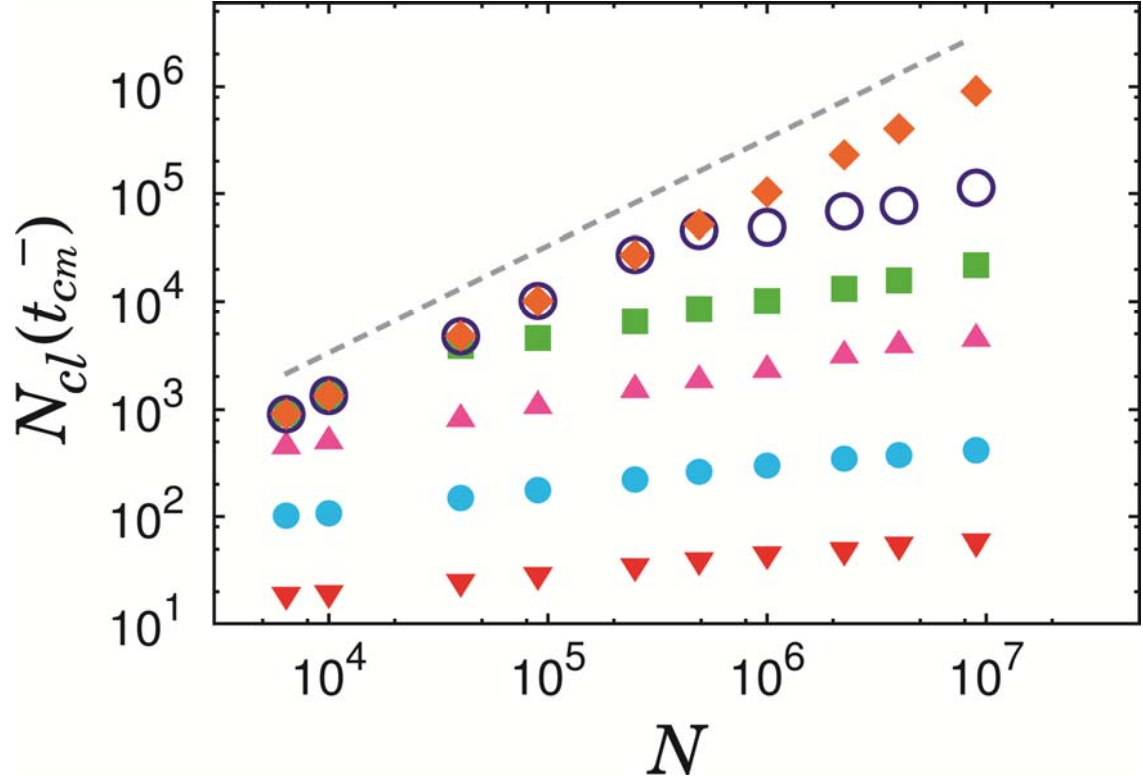

**Figure S11.** Plot of the number of clusters $N_{cl}$ at $t_{cm}^{-}$ as a function of system size $N$ for several $w$ values for the Gaussian model in two dimensions. Data are for $w = 0, 10^{-8}, 10^{-6}, 10^{-4}, 10^{-2}$, and 1 from the top. Dashed line is a guideline with slope 1. For $w = 10^{-8}$ (open circle), a crossover can be seen from the behavior $N_{cl} \sim N$ in small-$N$ region to the one $N_{cl} \sim N^{\alpha}$ with $\alpha < 1$ in large-$N$ region. All data are obtained after averaging over $10^3$ configurations.

**Table S1.** List of numerical values of $m_c(d)$ for dimensions $d = 2-6$. The values for $\zeta$ and $d_{BB}$ were measured in (*16*). * The standard deviation becomes larger than the mean value $m_c$.

| $d$ | $\zeta$ | $d_{BB}$ | $m_c$ |
|---|---|---|---|
| 2 | $0.50 \pm 0.03$ | $1.215 \pm 0.002$ | $2.55 \pm 0.01$ |
| 3 | $1.0 \pm 0.1$ | $2.498 \pm 0.005$ | $5.98 \pm 0.07$ |
| 4 | $1.3 \pm 0.5$ | $3.74 \pm 0.08$ | $16.99 \pm 5.23$ |
| 5 | $1.4 \pm 0.6$ | $4.9 \pm 0.2$ | $50^*$ |
| 6 | $1.5 \pm 0.7$ | $6.0 \pm 0.1$ | $\infty$ |